\documentclass{article}
\usepackage{arxiv}
\usepackage[utf8]{inputenc} 
\usepackage[T1]{fontenc}    
\usepackage{hyperref}       
\usepackage{url}            
\usepackage{booktabs}       
\usepackage{amsfonts}       
\usepackage{nicefrac}       
\usepackage{microtype}      
\usepackage{lipsum}
\usepackage{authblk}
\usepackage[misc]{ifsym}
\usepackage{graphicx}
\usepackage{multicol}
\usepackage{multirow}
\usepackage{mathtools,amssymb,amsmath}
\usepackage{algorithm}
\usepackage{algorithmic}
\usepackage{bbm}
\usepackage{comment}
\usepackage{color}

\DeclareMathOperator{\tat}{T}
\DeclareMathOperator{\cat}{cat}

\DeclareMathOperator{\wti}{WTI}

\DeclareMathOperator{\sign}{sign}
\usepackage{newfloat}
\usepackage{listings}
\DeclareMathOperator{\Encoder}{Encoder}
\DeclareMathOperator{\MHA}{MHA}

\DeclareMathOperator{\LN}{LN}
\DeclareMathOperator{\FFN}{FFN}
\def\ie{\emph{i.e.}}
\def\eg{\emph{e.g.}}
\def\etal{\emph{et al.}}
\begin{document}
\title{M2HF: Multi-level Multi-modal Hybrid Fusion for Text-Video Retrieval}
\author[1]{Shuo Liu\thanks{This work was done when Shuo Liu was interned at Tencent.}}
\author[1]{Weize Quan}
\author[2]{Ming Zhou}
\author[3]{Sihong Chen\thanks{whalechen@tencent.com}}
\author[3]{Jian Kang}
\author[3]{Zhe Zhao}
\author[3]{CHEN CHEN}
\author[1]{Dong-Ming Yan\thanks{yandongming@gmail.com}}
\affil[1]{NLPR, Institute of Automation, Chinese Academy of Sciences, Beijing, China}
\affil[2]{Donghua University, Shanghai, China}
\affil[3]{Tencent TEG AI, Shenzhen, China}
\renewcommand\Authands{ and }
\maketitle
\begin{abstract}
Videos contain multi-modal content, and exploring multi-level cross-modal interactions with natural language queries can provide great prominence to text-video retrieval task (TVR). However, new trending methods applying large-scale pre-trained model CLIP for TVR do not focus on multi-modal cues in videos. Furthermore, the traditional methods simply concatenating multi-modal features do not exploit fine-grained cross-modal information in videos. In this paper, we propose a multi-level multi-modal hybrid fusion (M2HF) network to explore comprehensive interactions between text queries and each modality content in videos. Specifically, M2HF first utilizes visual features extracted by CLIP to early fuse with audio and motion features extracted from videos, obtaining audio-visual fusion features and motion-visual fusion features respectively. Multi-modal alignment problem is also considered in this process. Then, visual features, audio-visual fusion features, motion-visual fusion features, and texts extracted from videos establish cross-modal relationships with caption queries in a multi-level way. Finally, the retrieval outputs from all levels are late fused to obtain final text-video retrieval results. Our framework provides two kinds of training strategies, including an ensemble manner and an end-to-end manner. Moreover, a novel multi-modal balance loss function is proposed to balance the contributions of each modality for efficient end-to-end training. M2HF allows us to obtain state-of-the-art results on various benchmarks, \eg, Rank@1 of 64.9\%, 68.2\%, 33.2\%, 57.1\%, 57.8\% on MSR-VTT, MSVD, LSMDC, DiDeMo, and ActivityNet, respectively.    
\end{abstract}


\section{Introduction}

With billions of videos uploaded at any time on online video platforms, it is worthwhile to retrieve the best corresponding video for a given query to efficiently access the desired video. Therefore, the tasks of Text-to-Video (T2V) and Video-to-Text (V2T) are tackled in this paper. T2V aims to obtain the ranking of all candidate videos for each caption query, while V2T finds the ranking of all candidate captions for each video query. 

Unlike images, video is a kind of media owning multiple different modalities. Therefore, considering and exploring different modalities in videos is really necessary for video understanding. Some traditional methods~\cite{Gabeur2020multi,dzabraev2021mdmmt} have paid attention to this point. For example, MMT~\cite{Gabeur2020multi} first extracted audio, visual, motion, face, scene, appearance, and ocr multi-modal features for obtaining better video representation. However, the coarse-grained way of concatenating all these features and feeding them into a transformer encoder may lead to models wishing to focus on certain modalities, while other informative modalities are overwhelmed and ignored, hindering the full video understanding of multi-modal contents. 

Recently, some methods~\cite{luo2021clip4clip,Cao2022visual,min2022hunyuan_tvr} have tried to utilize the pre-trained text-image retrieval model CLIP (contrastive language-image pretraining)~\cite{clip}, which is trained on 400 million text-image pairs to learn representation between text and image, as the backbone to conduct text-video retrieval task. For instance, CLIP4Clip~\cite{luo2021clip4clip} first utilized CLIP to extract the visual frame features and the caption token features, and then accumulated the similarity scores between frame-level video features and sentence-level caption features for the final results, achieving a notable improvement in TVR. Based on CLIP4Clip, HunYuan\_tvr~\cite{min2022hunyuan_tvr} formulated video-sentence, clip-phrase, and frame-word relationships to explore hierarchical cross-modal interactions. Unfortunately, these CLIP-based works entirely ignore other rich multi-modal signals, such as audio, motion, and text in videos. 

In this paper, we propose a novel method, Multi-Level Multi-Modal Hybrid Fusion (M2HF), which not only completes multi-modal contents in a fine-grained multi-level way, but also embraces the powerful pre-trianed model CLIP. First, M2HF early fuses audio and motion features respectively with visual features extracted by CLIP, producing audio-guided visual features and motion-guided visual features, which explicitly pay attention to sound source and moving objects. Then, we exploit the relationships between visual features, audio-guided visual features, motion-guided visual features, and text contents from ASR (automatic speech recognition) in a multi-level way with caption queries. When encountering the modality missing, we present a simple alignment strategy to align it. Finally, the results at each level are integrated by selecting the best ranking for each text-video pair as the final retrieval result.




Our contributions can be summarized as follows: 
\begin{itemize}
\item We propose a \textit{Multi-Level Multi-Modal Hybrid Fusion} network which improves the performance of text-video retrieval task by utilizing the capability of CLIP and exploring rich multi-modal information. 

\item We explore \textit{multi-modal complement and multi-modal alignment} based on early fusion strategy. Multi-level framework is designed by building relationships between language queries and visual, audio, motion, and text information extracted from videos for fully exploiting interactions between caption and video.

\item We devise a \textit{late multi-modal balance fusion} method to integrate results of each level by choosing the best ranking among them for obtaining the final ranking result. 

\item We introduce a novel \textit{multi-modal balance loss} for end-to-end training (E2E) of TVR task. M2HF can also be trained in an ensemble manner. The experimental results on the public datasets show remarkable performance in E2E and ensemble settings. 

\item \textit{M2HF} achieves new state-of-the-art Rank@1
retrieval results of 64.9\%, 68.2\%, 33.2\%, 57.1\%, 57.8\% on MSR-VTT, MSVD, LSMDC, DiDemo, and ActivityNet.
\end{itemize}

\section{Related Work}

\subsection{Multi-modal Fusion}

\subsubsection{Early Fusion.} Such methods mainly fuse multiple modalities at the feature level. Bilinear pooling-based approaches fuse two modalities by learning a joint representation space, \eg, MLB (low-rank bilinear pooling)~\cite{Kim2017Hadamard} and MFB (multi-modal factorized bilinear pooling)~\cite{Yu2017Multimodal}, etc. Attention-based methods fuse features from different modalities based on the correlation, including channel-wise attention~\cite{hu2018squeeze}, non-local model~\cite{Wang2018Nonlocal}, and transformer~\cite{Vaswani2017Attention,Xu2020Cross}, etc. 


\subsubsection{Late Fusion.}
Late fusion, also known as decision-level fusion, first trains different models on different modalities, and then fuses the predictions of these trained models~\cite{Snoek2005Early}. Late fusion methods mainly design the different combination strategies to merge models' outputs, such as voting, average combination, ensemble learning, and other combination methods.

\subsubsection{Hybrid Fusion.} The hybrid fusion method combines early fusion and late fusion, absorbing the advantages of both fusions. In this work, we employ a hybrid fusion mechanism to simultaneously exploit cross-modal relationships and tolerate the intervention and asynchrony of different modalities.


\subsection{Text-Video Retrieval}

For TVR, two research directions mainly exist: multi-modal features and large-scale pre-trained models.

\subsubsection{Text-Video Retrieval based on Multi-modal Features.}
One direction applies rich multi-modal cues to retrieve videos. MMT~\cite{Gabeur2020multi} encoded seven modalities such as audio, visual, and motion separately, and then fed them into a transformer for better video representation. 
MDMMT~\cite{dzabraev2021mdmmt} extends MMT by optimizing on training datasets.
MDMMT-2~\cite{kunitsyn2022mdmmt} introduced a three-stage training process and double positional encoding for better retrieval performance. However, these methods mainly input various multi-modal features into an encoder producing video representations. This fusion method is somewhat simple and coarse-grained. Instead, our method can utilize a fine-grained hybrid fusion method to fuse multi-modal features in a multi-level manner.


\begin{figure*}[t]
\centering
\includegraphics[width=0.89\linewidth]{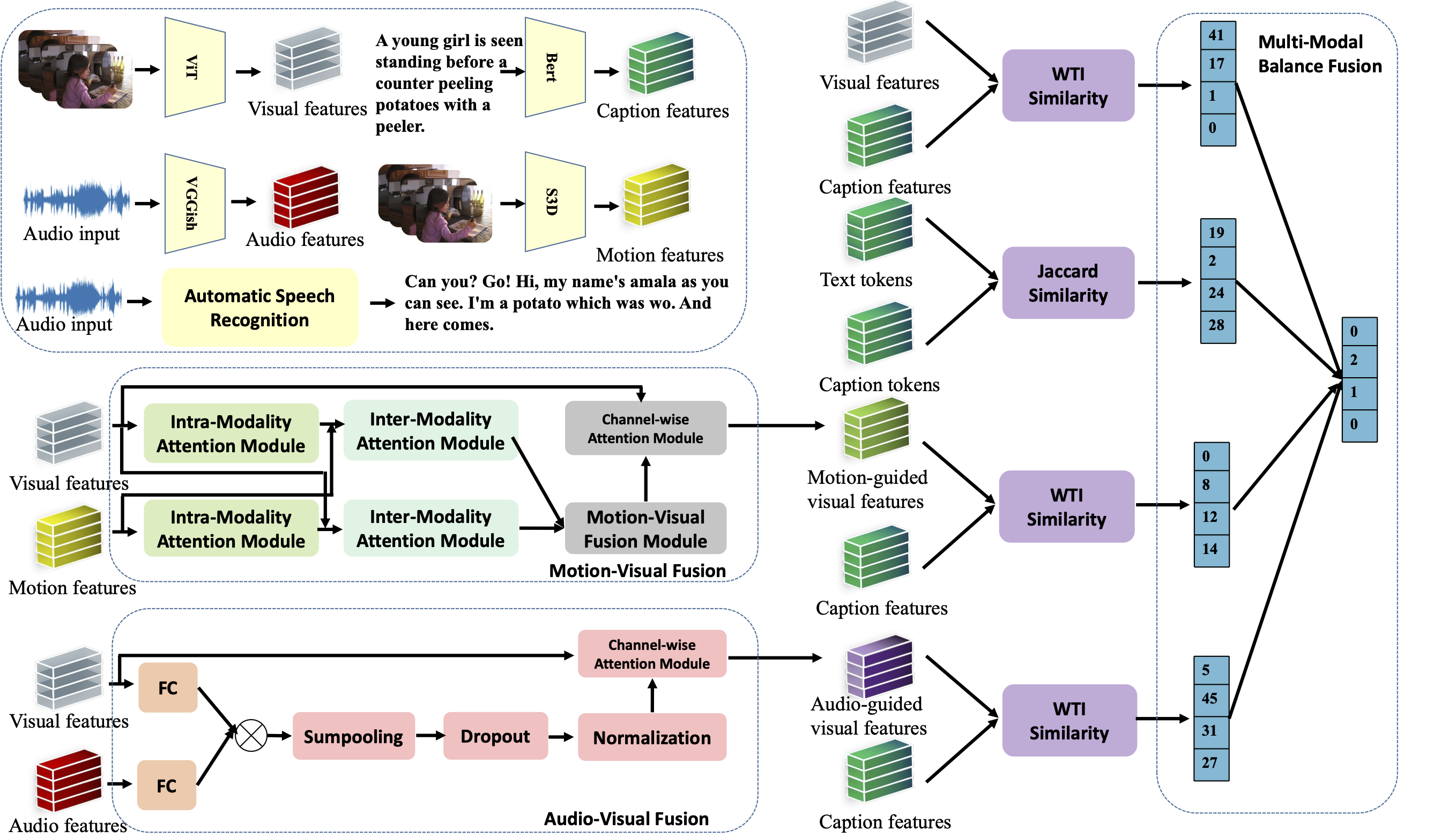}
\caption{The architecture of our multi-level multi-modal hybrid fusion network (M2HF) for text-video retrieval.}
\label{fig:Main_Network}
\end{figure*}


\subsubsection{Text-Video Retrieval based on CLIP.}
Another direction attempts to utilize pre-trained CLIP~\cite{clip} as the backbone for TVR task. The seminal work CLIP4Clip~\cite{luo2021clip4clip} exploited CLIP to extract features of visual frames and captions, and then computed the similarity scores between video and caption features. Based on CLIP, Fang \etal~\cite{fang2021clip2video} introduced temporal difference block and temporal alignment block to enhance temporal relationships between video frames and video captions. Cheng \etal~\cite{cheng2021improving} proposed a novel dual Softmax loss (DSL). Wang \etal~\cite{wang2022disentangled} carefully studied the cross-modality interaction process and representation learning for TVR, and proposed a disentangled framework, including a weighted token-wise interaction (WTI) block and a channel decorrelation regularization block, to model the sequential and hierarchical representation. Very recently, Gorti \etal~\cite{Gorti2022xpool} leveraged CLIP as a backbone and proposed a parametric text-conditioned pooling to aggregate video frame representations based on the similarity between video frame and text. However, these CLIP-based methods mainly focus on the visual modality, while ignoring other information in videos, such as motion, audio, and text, which are still important cues for TVR task.  




\section{Proposed Method}
\subsection{Overall Architecture}
Fig.~\ref{fig:Main_Network} illustrates the entire pipeline of our M2HF. Given a set of video-caption pairs $\{(V_{1},c_{1}),...,(V_{n},c_{n})\}$, our method measures the similarity of video and caption from four levels. To realize multi-modal complement, M2HF establishes relationships and conducts similarity computation between caption $c_{j}$ and visual $v_{i}$, audio $a_{i}$, motion $m_{i}$, and text $t_{i}$ extracted from video $V_{i}$, respectively. 

Multi-modal fusion is in a hybrid fusion way, where motion and audio features are early fused with visual features, \ie, motion-visual fusion and audio-visual fusion in Fig.~\ref{fig:Main_Network}, and all levels' ranking results are late fused for the final retrieval results by selecting the best ranking in the output of each level. We aggregate multi-modal cues in a fine-grained way for more accurate retrieval performance, and provide a multi-modal alignment method for the situation of modality missing.
Furthermore, two training strategies (E2E and Ensemble) are provided in this paper, and a novel multi-modal balance loss is proposed to serve E2E training by minimizing each pair score and calculating the balanced loss.

\subsection{Visual-to-Caption Level}
Visual-to-caption level is designed for making cross-modality relationship between visual features from video and caption query features. Image encoder ViT~\cite{dosovitskiy2020vit} and text encoder Bert~\cite{Devlin2019Bert} of CLIP is first used to extract visual features ($v_{i} \in \mathbb{R}^{F \times d_{v}}$) and caption features ($c_{i} \in \mathbb{R}^{T \times d_{c}}$), respectively, where $F$ is the number of video frames, $T$ is the number of caption tokens,  $d_{v}$ and $d_{c}$ represents the dimensions of visual and caption features respectively. To compute the similarity matrix $\mathcal{S}_{c-v}$ between caption features and visual features, we choose the weighted token-wise interaction (WTI) function~\cite{wang2022disentangled}. The entire process is computed as:
\begin{equation}
\begin{split}
c2v\_logits & = \sum_{p=1}^{T}f^{p}_{cw,\theta}(c_{i})max^{L_{v}}_{q=1}(\frac{c^{p}_{i}}{\left \|c^{p}_{i}\right \|_{2}})^{\tat}{v^{q}_{i}}, \\
v2c\_logits & = \sum_{q=1}^{F}f^{q}_{vw,\theta}(v_{i})max^{L_{c}}_{p=1}(\frac{c^{p}_{i}}{\left \|c^{p}_{i}\right \|_{2}})^{\tat}(\frac{v^{q}_{i}}{\left \|v^{q}_{i}\right \|_{2}}), \\
\mathcal{S}_{c-v}[i,i]  & = \wti(c_{i},v_{i}) = \frac{c2v\_logits+ v2c\_logits}{2.0},
\end{split}
\label{equ:wti}
\end{equation}
where $f_{cw,\theta}$ and $f_{vw,\theta}$ both are the combination of the MLP (multilayer perceptron) and a Softmax, $i$ is a sampled index, $p$ and $q$ denote the index of caption token and video frame. 


\subsection{Audio-to-Caption Level}
\label{subsec:audio-to-caption}
In the audio-to-caption level, audio features and visual features are early fused to highlight the visual semantic information related to audio. Then, the audio-guided visual features are used to build connections with caption features. Audio features ($a_{i} \in \mathbb{R}^{F \times d_{a}}$) are extracted from the log mel-spectrogram via the VGGish~\cite{Shawn2017CNN} pre-trained on AudioSet~\cite{Gemmeke2017Audio}, where $d_{a}$ is the dimension of audio features. 
We adopt MFB-based method in text-to-video task to early fuse audio and visual features, yielding high-level semantic audio-visual fusion features $\mathcal{F}_{av_{i}} \in \mathbb{R}^{F \times d_{v}}$.  Specifically, audio features $a_{i}$ and visual features $v_{i}$ are projected and aligned as the same dimension $kd$ using linear layers and ReLU. The aligned audio and visual features are multiplied and fed into sum pooling layer with the kernel size $k$. The formulation is as follows:     
\begin{equation}
\mathcal{F}_{av_{i}} = \text{Drop}(\text{SP}(\Psi^{\tat}a_{i} \odot \Phi^{\tat}v_{i},k)),
\end{equation}
where $\Psi \in \mathbb{R} ^{d_{a} \times(kd)}$ and $\Phi \in\mathbb{R}^{d_{v} \times(kd)}$ are two learnable matrices, $\odot$ represents element-wise product, $\text{SP}(\cdot,k)$ is the sum pooling with kernel size $k$ and stride $k$, and $\text{Drop}(\cdot)$ is a dropout layer to prevent the over-fitting. To stabilize the model training, power and $L_{2}$ normalizations are utilized:
\begin{equation}
\label{equ:map_mfb_s}
\mathcal{F}_{av_{i}} \leftarrow \sign (\mathcal{F}_{av_{i}}) \sqrt{\left | \mathcal{F}_{av_{i}} \right |}, 
\mathcal{F}_{av_{i}} \leftarrow \mathcal{F}_{av_{i}}/\left \| \mathcal{F}_{av_{i}} \right \|.
\end{equation} 

Next, the audio-visual fusion feature guides the raw visual features $v_{i}$ by channel-wise attention operation for obtaining the final audio-guided visual features. A squeeze-and-excitation operation~\cite{hu2018squeeze} is applied to produce channel-wise attentive weights ($\mathcal{W}_{i}^{\mathcal{A}} \in \mathbb{R}^{d_{v}\times 1}$). This process is formulated as:
\begin{equation}
\mathcal{W}_{i}^{\mathcal{A}} = \delta (\mathbf{W}_{2} \sigma  (\mathbf{W}_{1}(\mathcal{F}_{av_{i}}))),
\end{equation}
where $\mathbf{W}_{1} \in \mathbb{R}^{d_v \times d}$ and $\mathbf{W}_{2} \in \mathbb{R}^{d \times d_v}$ are two linear transformations with $d=\frac{d_v}{2}$; $\sigma$ and $\delta$ denote the ReLU and sigmoid functions, respectively.

The final audio-guided visual features are obtained via:
\begin{equation}
av_{i} = \mathcal{W}_{i}^{\mathcal{A}} \odot v_{i}.
\end{equation}

Similar to visual-to-caption level, the relationship between audio-guided visual features and caption features is formulated by WTI. The detailed formula of the similarity matrix $\mathcal{S}_{c-a}$ is similar with Eq.(\ref{equ:wti}) replacing $v_{i}$ with $av_{i}$.

Considering that not all videos exist audio signals, \ie, modality missing problem, we pad missing audio features with element $1$. The primary idea of this alignment strategy is that the guidance mechanism can still work by guiding with original visual features.   

\subsection{Motion-to-Caption Level}
Motion-to-caption level is proposed to early fuse motion features with visual features obtaining motion-guided visual features, which explicitly considers the object movement in visual modality. The fused features then compute the similarity with caption features. In our work, S3D~\cite{Xie2018RethinkingSF} pre-trained on the Kinetics~\cite{carreira2017quo} is applied to extract motion features ($m_{i} \in \mathbb{R}^{F \times d_{m}}$), where $d_{m}$ is the dimension of motion features.

For the fusion of motion features and visual features, we utilize the encoder of transformer block~\cite{Vaswani2017Attention}. The detailed calculation process is as follows:
\begin{equation}
\begin{split}
&\Encoder(Q,K,V) = \LN(X+Y),\\
&X=\MHA(\tilde{Q},\tilde{K},\tilde{V}), Y=\FFN(\LN(X+\tilde{Q})),\\
&\tilde{Q}=Q\mathbf{W}_{\tilde{Q}}, \tilde{K}=K\mathbf{W}_{\tilde{K}}, \tilde{V}=V\mathbf{W}_{\tilde{V}},
\end{split}
\label{equ:Encoder}
\end{equation}
where $Q$, $K$, $V \in \mathbb{R}^{F \times d}$ are input features of transformer's encoder; $\mathbf{W}_{\tilde{Q}},\mathbf{W}_{\tilde{K}},\mathbf{W}_{\tilde{V}} \in \mathbb{R}^{d \times d}$ are projection matrices; $\LN$ refers to the layer normalization; $\MHA$ is the multi-head attention~\cite{Vaswani2017Attention} with 4 heads; and $\FFN$ is the feed forward network. 

As shown in Fig.~\ref{fig:Main_Network}, motion features $m_{i}$ and visual features $v_{i}$ are first fed into the intra-modality attention module to learn the informative segments of each modality. 
The motion modality is taken as an example to explain the intra-modality attention module. Specifically, motion features are first projected yielding query features ($Q \in \mathbb{R}^{F \times d_{m}}$), key features ($K \in \mathbb{R}^{F \times d_{m}}$), and value features ($V \in \mathbb{R}^{F \times d_{m}}$). They are then fed into encoder of transformer producing the self-attentive motion features $m^{self} = \Encoder(Q,K,V)$ via Eq.~\ref{equ:Encoder}. Self-attentive visual features $v^{self}$ are obtained using the same way. 

Next, inter-modality attention module is introduced to exploit relationship between motion and visual features via encoder of transformer as well. Different from the intra-modality computation, key and value features of the inter-modality attention are the concatenation of one modality features and the self-attentive features of another modality. Cross-modality features $m^{cross}$ and $v^{cross}$ are obtained as:
\begin{equation}
\begin{split}
&m^{cross} = \Encoder(m_{i},\cat(m_{i},v^{self}_{i}),\cat(m_{i},v^{self}_{i})),\\
&v^{cross} = \Encoder(v_{i},\cat(v_{i},m^{self}_{i}),\cat(v_{i},m^{self}_{i})),
\end{split}
\end{equation}
where $\cat$ is the concatenation of two features in temporal dimension. The cross-modality features are then integrated by the encoder of transformer yielding motion-visual fusion features $F_{mv_{i}} \in \mathbb{R}^{F\times d_{v}}$ via:
\begin{equation}
\begin{split}
&F_{mv_{i}} = \Encoder(Q,K,V),\\
&Q = m^{cross}_{i} \odot v^{cross}_{i}, K,V = \cat(m^{cross}_{i},  v^{cross}_{i}).
\end{split}
\end{equation} 

After that, $F_{mv_{i}}$ is used to guide the visual features to highlight the moving objects. The guidance weights are first estimated via the squeeze-and-excitation block~\cite{hu2018squeeze} as follows:
\begin{equation}
\mathcal{W}_{i}^{\mathcal{M}} = \delta (\mathbf{W}_{4}\sigma(\mathbf{W}_{3}(\mathcal{F}_{mv_{i}}))),
\end{equation}
where $\mathbf{W}_{3} \in \mathbb{R}^{d_v \times d}$ and $\mathbf{W}_{4} \in \mathbb{R}^{d \times d_v}$ are two linear transformations with $d=\frac{d_v}{2}$.
Motion-guided visual features $mv_{i}$ are achieved via:
\begin{equation}
mv_{i} = \mathcal{W}_{i}^{\mathcal{M}} \odot v_{i},
\end{equation}

Finally, the similarity matrix $\mathcal{S}_{c-m}$ between motion-guided visual features and caption features is calculated via the WTI, \ie, replacing $v_{i}$ in Eq.(\ref{equ:wti}) with $mv_{i}$. 


\subsection{Text-to-Caption Level}
Text information is extracted from video via ASR technology. The same modality can directly compute the similarity matrix $\mathcal{S}_{c-t}$ without intervention from other modalities. Jaccard scores~\cite{Jaccard1912Distrbution}  are formulated for each pair of caption and text. Before the formulation, several pre-processing operations are conducted. First, stop words including pronouns, integrated nouns, and other less representative words are filtered from text and caption. Then, the remaining words will be filtered again to keep only nouns since nouns are more representative than verbs, adverbs, prepositions, and others. Next, the filtered tokens are converted to the same root. Finally, all the letters lowercase yielding the final set of text $S_{t}$ and caption $S_{c}$. The calculation of Jaccard score is as:
\begin{equation}
\mathcal{S}_{c-t}[i,i]  = \text{Jaccard}(S_{c},S_{t}) = \frac{\text{len}(S_{c}\cap S_{t})}{\text{len}(S_{c} \cup S_{t})},
\end{equation}
where $\text{Jaccard}(\cdot, \cdot)$ computes the jaccard correlation, and $\text{len}(\cdot)$ calculates the number of each set's tokens.

\subsection{Ensemble and E2E Text-to-Video Retrieval}
\subsubsection{Ensemble Retrieval.} Inspired by ensemble learning, we
first train each multi-modality level model independently, and then fuse their predictions. Dual softmax loss (DSL)~\cite{cheng2021improving} is applied for visual-to-caption level, audio-to-caption level, and motion-to-caption level. DSL pursues the dual optimal match and thus obtain the good retrieval performance. Specifically, the similarity matrices $\mathcal{S}_{c-v}$, $\mathcal{S}_{c-a}$, and $\mathcal{S}_{c-m}$ are fed into DSL function. 
We take $\mathcal{S}_{c-v}$ as an example, the loss of visual-to-caption level ($\mathcal{L}_{v} = -\frac{1}{B}\sum_{i}^{B}\mathbf{L}_{v}$) formulates as follows:
\begin{equation}
\begin{split}
&P_{v2c}[i,j]  = \frac{e^{(\lambda\mathcal{S}_{c-v}[i,i])}}{\sum_{j=1}^{B}{e^{(\lambda\mathcal{S}_{c-v}[j,i])}}},  \\
&P_{c2v}[i,j]  = \frac{e^{(\lambda\mathcal{S}_{c-v}[i,i])}}{\sum_{j=1}^{B}{e^{(\lambda\mathcal{S}_{c-v}[i,j])}}}, \\
&\mathbf{L}_{v2c}[i]  = log(\frac{e^{(\eta \mathcal{S}_{c-v}[i,i] P_{v2c}[i,i])}}{\sum_{j=1}^{B}e^{(\eta \mathcal{S}_{c-v}[i,j] P_{v2c}[i,j])}}),\\
&\mathbf{L}_{c2v}[i]  =log(\frac{e^{(\eta \mathcal{S}_{c-v}[i,i] P_{c2v}[i,i])}}{\sum_{j=1}^{B}e^{(\eta \mathcal{S}_{c-v}[j,i] P_{c2v}[j,i])}}), \\
&\mathbf{L}_{v}  = \mathbf{L}_{v2c} + \mathbf{L}_{c2v},
\end{split}
\end{equation}
where $\lambda$ is a temperature hyper-parameter to smooth the gradients, $B$ is batch size, and $\eta$ is a logit scaling factor. $\mathcal{L}_{a}$ and $\mathcal{L}_{m}$ are obtained in the same way.
 
For the evaluation, a novel late fusion strategy, called multi-modal balance fusion (MMBF), is proposed to fuse the outputs of all levels by selecting the best ranking from each level. The ranking of each level is denoted as $\mathcal{R}_{c-v}$, $\mathcal{R}_{c-a}$, $\mathcal{R}_{c-m}$, and $\mathcal{R}_{c-t}$, which are obtained based on the respective similarity matrices. Then, the final ranking is   
\begin{equation}
\mathcal{R} = min(\mathcal{R}_{c-v}, \mathcal{R}_{c-a}, \mathcal{R}_{c-m}, \mathcal{R}_{c-t}),
\end{equation}
where $min$ is element-wise minimizing operation.

\subsubsection{E2E Retrieval.} In addition, we introduce a novel multi-modal balance loss (MMBL) to train the model in an end-to-end manner. Specifically, MMBL uses the minimum value of each level yielding the final balanced loss as follows:
\begin{equation}
\label{equ:mmbl}
\mathcal{L} =-\frac{1}{B}\sum_{i}^{B}min(\mathbf{{L}_{v}}, \mathbf{{L}_{a}}, \mathbf{{L}_{m}}).
\end{equation}
We also try other similar fusion methods, including average, element-wise maximizing, and element-wise adding, and find that the element-wise minimizing achieves the best performance. The evaluation of E2E Retrieval also uses MMBF.

\begin{figure*}[t]
\centering
\includegraphics[width=\linewidth]{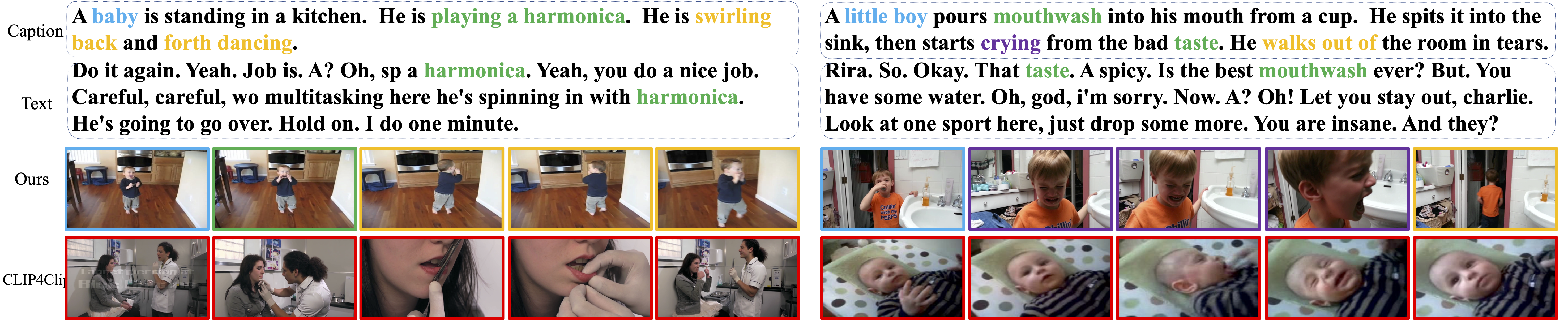}
\caption{Qualitative comparisons of our method with CLIP4Clip~\cite{luo2021clip4clip}. ``Blue'', ``Green'', ``Orange'', and ``Purple'' represents visual, text, motion, and audio cues, respectively.}
\label{fig:Qualitative_results}
\end{figure*}

\begin{table*}[htp]
\centering
\begin{tabular}{|l|l l l l l|l l l l l|}
    \hline
    \multirow{2}{*}{Methods} & & & T2V & & & & & V2T & & \\
     & R@1 & R@5 & R@10 & MdR & MnR & R@1 & R@5 & R@10 & MdR & MnR \\
    \hline \hline
    T2VLAD \cite{wang2021t2vlad} & 29.5 & 59.0 & 70.1 & 4.0 & - & - & - & - & - & - \\
    CLIP4Clip \cite{luo2021clip4clip} & 44.5 & 71.4 & 81.6 & 2.0 & 15.3 & 42.7 & 70.9 & 80.6 & 2.0 & - \\
    VCM \cite{Cao2022visual} & 43.8 & 71.0 & 80.9 & 2.0 & 14.3 & 45.1 & 72.3 & 82.3 & 2.0 & 10.7 \\
    X-Pool \cite{Gorti2022xpool} & 46.9 & 72.8 & 82.2 & 2.0 & 14.3 & - & - & - & - & - \\
    CAMOE \cite{cheng2021improving} & 48.8 & 75.6 & 85.3 & 2.0 & 12.4 & 50.3 & 74.6 & 83.8 & 2.0 & 9.9 \\
    DCR \cite{wang2022disentangled} & 55.3 & 80.3 & 87.6 & 1.0 & - & 56.2 & 79.9 & 87.4 & 1.0 & - \\
    Hun Yuan\_tvr (ViT-B/16) \cite{min2022hunyuan_tvr} & 55.0 & 80.4 & 86.8 & 1.0 & 10.3 & 55.5 & 78.4 & 85.8 & 1.0 & 7.7 \\
    Hun Yuan\_tvr (ViT-L/14) \cite{min2022hunyuan_tvr} & 53.2 & 77.6 & 83.9 & 1.0 & 10.1 & 54.0 & 78.8 & 87.1 & 1.0 & 8.3 \\    
    \hline
    Ours\_E2E (ViT-B/16) & 60.5 & 83.5 & 90.3 & 1.0 & \textbf{7.0} & 60.7 & 85.3 & 91.4 & 1.0 & 5.9 \\
    Ours\_E2E (ViT-L/14) & 60.9 & 83.6 & 89.6 & 1.0 & 8.3 & 60.8 & 83.6 & 90.2 & 1.0 & 5.7 \\
    Ours\_Ensemble (ViT-B/16) & 63.0 & 84.6 & 90.9 & 1.0 & 7.2 & 63.0 & 85.1 & 91.5 & 1.0 & 4.8 \\
    Ours\_Ensemble (ViT-L/14) & \textbf{64.9} & \textbf{85.0} & \textbf{91.6} & \textbf{1.0} & 7.9 & \textbf{66.2} & \textbf{85.8} & \textbf{91.9} & \textbf{1.0} & \textbf{5.0} \\    
    \hline    
\end{tabular}
\vskip -0.2cm 
\caption{Retrieval results on MSR-VTT 1K dataset.}
\label{tab:compar_msrvtt}
\end{table*}

\begin{figure*}[t]
\centering
\includegraphics[width=\linewidth]{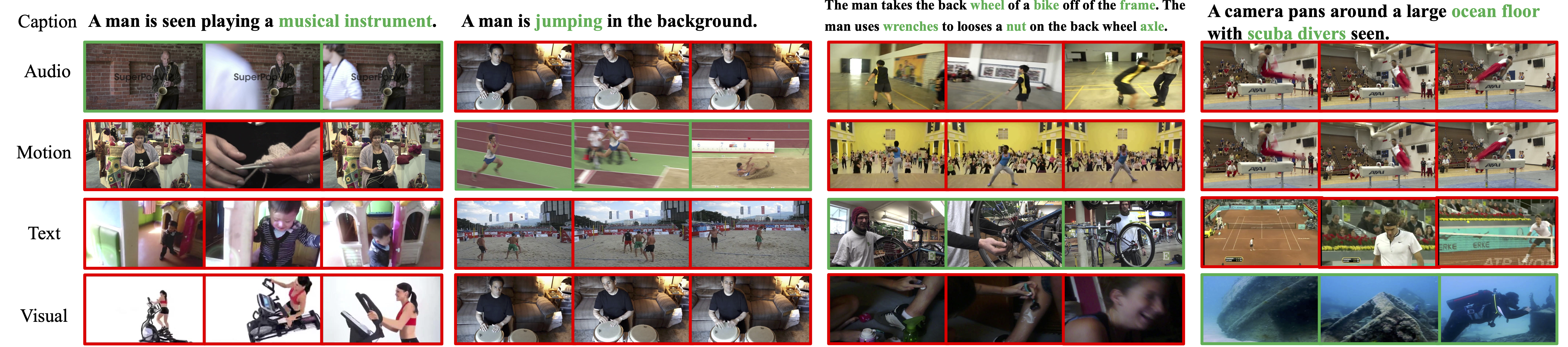}
\caption{Ablation studies of M2HF. ``Audio'', ``Motion'', ``Text'', and ``Visual'' are the model of only using corresponding level.}
\label{fig:Ablation_results}
\end{figure*}

\begin{table*}[htp]
\centering
\begin{tabular}{|l|l l l l l|l l l l l|}
    \hline
    \multirow{2}{*}{Methods} & & & T2V & & & & & V2T & & \\
     & R@1 & R@5 & R@10 & MdR & MnR & R@1 & R@5 & R@10 & MdR & MnR \\
    \hline \hline
    CLIP4Clip \cite{luo2021clip4clip} & 46.2 & 76.1 & 84.6 & 2.0 & 10.0 & 48.4 & 70.3 & 77.2 & 2.0 & - \\
    X-Pool \cite{Gorti2022xpool} & 47.2 & 77.4 & 86.0 & 2.0 & 9.3 & - & - & - & - & - \\
    CAMOE \cite{cheng2021improving} & 49.8 & 79.2 & 87.0 & - & 9.4 & - & - & - & - & - \\
    DCR \cite{wang2022disentangled} & 50.0 & 81.5 & 89.5 & 2.0 & - & 58.7 & 92.5 & 95.6 & 1.0 & - \\
    Hun Yuan\_tvr (ViT-B/16) \cite{min2022hunyuan_tvr} & 54.6 & 82.4 & 89.6 & 1.0 & 8.0 & 58.0 & 85.4 & 89.6 & 1.0 & 5.5 \\
    Hun Yuan\_tvr (ViT-L/14) \cite{min2022hunyuan_tvr} & 57.8 & 83.3 & 89.6 & 1.0 & 7.8 & 63.4 & 88.1 & 92.6 & 1.0 & 3.3 \\    
    \hline
    Ours\_E2E (ViT-B/16) & 60.0 & 84.8 & 90.8 & 1.0 & 6.4 & 70.0 & \textbf{90.0} & 93.5 & 1.0 & 3.3 \\
    Ours\_E2E (ViT-L/14) & 62.6 & 85.9 & 91.5 & 1.0 & 5.7 & 71.3 & 89.6 & 95.8 & 1.0 & 2.9 \\
    Ours\_Ensemble (ViT-B/16) & 61.1 & 86.0 & 91.5 & 1.0 & 6.3 & 69.7 & 86.9 & 92.8 & 1.0 & 3.5 \\
    Ours\_Ensemble (ViT-L/14) & \textbf{68.2}  & \textbf{88.6} & \textbf{92.9} & \textbf{1.0} & \textbf{5.1} & \textbf{75.5} & 83.0 & \textbf{96.0} & \textbf{1.0} & \textbf{2.3} \\    
    \hline    
\end{tabular}
\vskip -0.2cm 
\caption{Retrieval results on MSVD dataset.}
\label{tab:compar_msvd}
\end{table*}

\begin{table*}[htp]
\centering
\begin{tabular}{|l|l l l l l|l l l l l|}
    \hline
    \multirow{2}{*}{Methods} & & & T2V & & & & & V2T & & \\
     & R@1 & R@5 & R@10 & MdR & MnR & R@1 & R@5 & R@10 & MdR & MnR \\
    \hline \hline
    CLIP4Clip \cite{luo2021clip4clip} & 11.2 & 26.9 & 34.8 & 25.3 & - & - & - & - & - & - \\
    T2VLAD \cite{wang2021t2vlad} & 14.3 & 32.4 & - & 16 & - & 14.2 & 33.5 & - & 17 & - \\
    X-Pool \cite{Gorti2022xpool} & 25.2 & 43.7 & 53.5 & 8.0 & 53.2 & - & - & - & - & - \\
    CAMOE \cite{cheng2021improving} & 25.9 & 46.1 & 53.7 & - & 54.4 & - & - & - & - & - \\
    DCR \cite{wang2022disentangled} & 26.5 & 47.6 & 56.8 & 7.0 & - & 27.0 & 45.7 & 55.4 & 8.0 & - \\
    Hun Yuan\_tvr (ViT-B/16) \cite{min2022hunyuan_tvr} & 26.3 & 46.1 & 54.1 & 7.0 & 55.3 & 27.1 & 46.6 & 54.5 & 7.0 & 45.7 \\
    Hun Yuan\_tvr (ViT-L/14) \cite{min2022hunyuan_tvr} & 29.7 & 46.4 & 55.4 & 7.0 & 56.4 & 30.1 & 47.5 & 55.4 & 7.0 & 48.9 \\    
    \hline
    Ours\_E2E (ViT-B/16) & 27.4 & 45.2 & 53.2 & 9.0 & 46.7 & 27.0 & 45.7 & 53.8 & 8.0 & 43.1 \\
    Ours\_E2E (ViT-L/14) & 31.4 & 49.1 & 58.5 & 6.0 & 44.4 & 30.5 & 48.6 & 59.6 & 6.0 & 38.1 \\
    Ours\_Ensemble (ViT-B/16) & 30.0 & 49.0 & 58.5 & 6.0 & 40.4 & 29.9 & 49.4 & 58.5 & 6.0 & 35.2 \\
    Ours\_Ensemble (ViT-L/14) & \textbf{33.2}  & \textbf{54.3} & \textbf{63.8} & \textbf{4.0} & \textbf{34.9} & \textbf{34.8} & \textbf{55.0} & \textbf{64.3} & \textbf{4.0} & \textbf{28.8} \\    
    \hline    
\end{tabular}
\vskip -0.2cm 
\caption{Retrieval results on LSMDC dataset.}
\label{tab:compar_lsmdc}
\end{table*}

\begin{table*}[htp]
\centering
\begin{tabular}{|l|l l l l l|l l l l l|}
    \hline
    \multirow{2}{*}{Methods} & & & T2V & & & & & V2T & & \\
     & R@1 & R@5 & R@10 & MdR & MnR & R@1 & R@5 & R@10 & MdR & MnR \\
    \hline \hline
    CLIP4Clip \cite{luo2021clip4clip} & 41.4 & 58.2 & 79.1 & 2.0 & - & 42.8 & 69.8 & 79.0 & 2.0 & - \\
    CAMOE \cite{cheng2021improving} & 43.8 & 71.4 & - & - & - & 45.5 & 71.2 & - & - & - \\
    DCR \cite{wang2022disentangled} & 49.0 & 76.5 & 84.5 & 2.0 & - & 49.9 & 75.4 & 83.3 & 2.0 & - \\
    Hun Yuan\_tvr (ViT-B/16) \cite{min2022hunyuan_tvr} & 52.1 & 78.2 & 85.7 & 1.0 & 11.1 & 54.8 & 79.9 & 87.2 & 1.0 & 7.1 \\
    Hun Yuan\_tvr (ViT-L/14) \cite{min2022hunyuan_tvr} & 49.5 & 73.7 & 81.6 & 2.0 & 14.8 & 50.3 & 76.5 & 83.7 & 1.0 & 10.4 \\    
    \hline
    Ours\_E2E (ViT-B/16) & 53.0 & 76.7 & 84.5 & 1.0 & 11.5 & 53.7 & 76.2 & 84.9 & 1.0 & 8.1 \\
    Ours\_E2E (ViT-L/14) & 54.1 & 76.9 & 85.5 & 1.0 & 11.1 & 53.5 & 77.6 & 86.0 & 1.0 & 8.3 \\
    Ours\_Ensemble (ViT-B/16) & 55.1 & 79.3 & 85.5 & 1.0 & 10.0 & 56.2 & 79.0 & 86.0 & 1.0 & 7.3 \\
    Ours\_Ensemble (ViT-L/14) & \textbf{57.1}  & \textbf{79.3} & \textbf{87.5} & \textbf{1.0} & \textbf{9.5} & \textbf{58.0} & \textbf{80.4} & \textbf{89.6} & \textbf{1.0} & \textbf{7.1} \\    
    \hline    
\end{tabular}
\vskip -0.2cm 
\caption{Retrieval results on DiDeMo dataset.}
\label{tab:compar_didemo}
\end{table*}

\begin{table*}[htp!]
\centering
\begin{tabular}{|l|l l l l l|l l l l l|}
    \hline
    \multirow{2}{*}{Methods} & & & T2V & & & & & V2T & & \\
     & R@1 & R@5 & R@10 & MdR & MnR & R@1 & R@5 & R@10 & MdR & MnR \\
    \hline
    w/o Audio & 60.6 & \textbf{83.9} & 89.6 & 1.0 & 8.9 & 60.4 & 83.1 & \textbf{90.3} & 1.0 & 6.8 \\
    w/o Motion & 58.5 & 81.1  & 88.2 & 1.0 & 9.9 & 57.9 & 82.1 & 90.1 & 1.0 & 7.0 \\
    w/o Text & 59.2 & 82.2 & 88.3 & 1.0 & 8.8 & 58.9 & 82.5 & 89.5 & 1.0 & 6.0 \\
    w/o Visual & 59.9 & 82.9 & 88.9 & 1.0 & 9.0 & 59.6 & 82.4 & 89.0 & 1.0 & 6.4 \\ 
    Ours\_E2E (ViT-L/14) & \textbf{60.9} & 83.6 & \textbf{89.6} & \textbf{1.0} & \textbf{8.3} & \textbf{60.8} & \textbf{83.6} & 90.2 & \textbf{1.0} & \textbf{5.7} \\
    \hline
    w/o Audio & 61.4 & 83.6 & 90.1 & 1.0 & 8.6 & 62.3 & 83.6 & 90.6 & 1.0 & 5.6 \\
    w/o Motion & 62.3 & 83.7  & 90.6 & 1.0 & 8.7 & 63.5 & 83.9 & 90.7 & 1.0 & 5.6 \\
    w/o Text & 63.4 & 83.9 & 90.5 & 1.0 & 8.3 & 64.9 & 84.5 & 91.0 & 1.0 & 5.4 \\
    w/o Visual & 62.1 & 82.2 & 90.2 & 1.0 & 9.0 & 62.9 & 83.9 & 90.7 & 1.0 & 5.8 \\    
    Ours\_Ensemble (ViT-L/14) & \textbf{64.9} & \textbf{85.0} & \textbf{91.6} & \textbf{1.0} & \textbf{7.9} & \textbf{66.2} & \textbf{85.8} & \textbf{91.9} & \textbf{1.0} & \textbf{5.0} \\  
    \hline    
\end{tabular}
\vskip -0.2cm
\caption{Ablation studies on MSR-VTT 1K dataset.}
\label{tab:msrvtt_ablation_studies}
\end{table*}

\section{Experiments}

\subsection{Experimental Settings}

\subsubsection{Datasets.}

We use five common benchmarks: MSR-VTT, MSVD, LSMDC, DiDeMo, and ActivityNet to conduct extensive experiments to validate our method.

\textbf{MSR-VTT}~\cite{xu2016msr} is a large-scale dataset containing 10,000 video clips and each video clip is described with 20 natural sentences via Amazon Mechanical Turks. Following the setting~\cite{yu2018joint}, 9,000 and 1,000 videos are used for training and testing, respectively.

\textbf{MSVD}~\cite{chen2011collecting} has 1,970 video clips, and each video clip contains about 40 sentences. We adopt the original data split, 1,200 videos for training, 100 videos for validation, and 670 videos for testing.

\textbf{LSMDC}~\cite{rohrbach2015long} is composed of 118,081 video clips extracted from 202 movies and each video clip has a caption. The validation set and evaluation set contains 7,408 and 1,000 videos, respectively. 

\textbf{ActivityNet}~\cite{krishna2017dense} 
contains 20,000 YouTube videos with 100k captions. Standard split, the training set has 10,009 videos and the validation set has 4,917 videos, is followed. Like~\cite{zhang2018cross}, we concatenate all the captions of a video as a paragraph.

\textbf{DiDeMo}~\cite{anne2017localizing} contains 10,000 videos annotated with 40,000 sentences. All captions of a video are concatenated into a paragraph~\cite{liu2019use}.

\subsubsection{Metrics.}
Following the standard retrieval metrics, Recall at rank N (R@N, higher is better), mean rank (MnR, lower is better), and median rank (MdR, lower is better).

\subsubsection{Implementation Details.}
Our method is implemented with PyTorch 1.7.1, and is trained on NVIDIA Tesla A100 GPU. We set the initial learning rate as $1e-7$ for the CLIP and $1e-4$ for the remaining parameters, respectively. For MSR-VTT, MSVD, and LSMDC, the frame number $F$ and token number $T$ are 12 and 32, respectively; For DiDeMo and ActivityNet, $F=64$ and $T=64$. Adam optimizer with batch size of 128 is used for training the model with 5 epoch.

\subsection{Comparison with State-of-the-art Methods}
In this subsection, we compare M2HF with state-of-the-art methods on MSR-VTT, MSVD, LSMDC, DiDeMo, and ActivityNet benchmarks. 

Table \ref{tab:compar_msrvtt} shows results of MSR-VTT, which can be seen that our M2HF significantly surpasses CLIP4Clip by 20.4\% R@1 and outperforms a very recent parallel work Hun Yuan\_tvr by 9.9\%. Table \ref{tab:compar_msvd} shows that M2HF achieves 10.4\% improvement on the MSVD compared to Hun Yuan\_tvr. For LSMDC as shown in Table \ref{tab:compar_lsmdc}, our approach obtains the gain over Hun Yuan\_tvr by 3.5\%. As reported in Table \ref{tab:compar_didemo}, M2HF remarkably outperforms the state-of-the-art method by 7.6\% for DiDeMo. All the quantitative results consistently illustrate the superiority of M2HF.

Fig.~\ref{fig:Qualitative_results} shows two qualitative comparison examples, which show that only using visual modality is not enough to represent videos well. In contrast, our multi-modal complement method provides multi-modal cues to obtain more accurate results. For the first example, images are helpful to match ``baby'' word. The harmonica sound made by this baby and the text information from off-screen sound can be associated with ``harmonica''. The baby's movements are matched with ``swirling back'' and ``forth dancing''. In the second example, ``little boy'' corresponds to a semantic visual target. The keywords ``mouthwash'' and ``taste'' in the text match the relevant tokens in caption. The ``crying'' sound made by this little boy is captured with the help of audio signals. His moving figures are related to ``walk out of''. However, visual-based method CLIP4clip can only retrieve ``a little boy''.

\subsection{Ablation Studies}
As reported in Table \ref{tab:msrvtt_ablation_studies}, ablation experiments on MSR-VTT are conducted to evaluate the effect of each level in M2HF. ``w/o Audio'', ``w/o Motion'', ``w/o Text'', and ``w/o Visual'' represents removing the relevant level from M2HF. Quantitative results demonstrate that each modality contributes to the performance of text-video retrieval. 

Fig.~\ref{fig:Ablation_results} shows the qualitative studies of our proposed method. ``Audio'', ``Motion'', ``Text'', and ``Visual'' are the effect of only using corresponding level for T2V retrieval. The green and red boxes represent the true and false retrieval results, respectively. These four examples explain the advantages of each modality. For the first column, ``Audio'' level can catch the instrument's sound source with the guidance of audio signal, however, other levels cannot provide the same contribution. The second example utilizes the motion features of jumping to predict the right retrieval, where ``Motion'' level can pay attention to the moving objects. The third one shows the effect of ``Text'' level, and there are six same keywords between caption tokens and text tokens, including ``wheel'', ``bike'', ``frame'', ``wrench'', ``nut'', and ``axle''. This level is really helpful for those cases containing lots of abstract nouns. For the last example, there are no sound and moving objects. Therefore, the ability of visual is powerful to detect ``ocean floor'' and ``scuba divers''. To this end, quantitative and qualitative results illustrate the superiority of our multi-level and multi-modal method for TVR.

In addition, when removing the multi-modal alignment (in Sec.~\ref{subsec:audio-to-caption}), the retrieval performance drops from 53.0\% to 52.5\% on DiDeMo. We also conduct the ablation experiments of multi-modal balance loss (Eq.~\eqref{equ:mmbl}) on MSR-VTT. Compared with element-wise minimizing (60.5\%), the R@1 metric of other fusion methods is lower, including average (55.7\%), element-wise maximizing (54.6\%), and element-wise adding(55.2\%). 


More detailed experimental results and ablation studies are included in the \emph{Supplementary Material}.


\section{Conclusion}

Based on the multi-modal nature of videos, in this paper, we proposed a novel multi-level multi-modal hybrid fusion network for text-video retrieval. The core idea is to explore fine-grained multi-modal cues in a multi-level way, and M2HF can also leverage the powerful knowledge from pre-trained text-image retrieval model (\ie, CLIP). To solve multi-modal complement and multi-modal alignment, we introduced a hybrid fusion methods. Moreover, two training strategies are exploited and implemented: end-to-end training with a multi-modal balance loss and ensemble training with a multi-modal balance fusion. Extensive quantitative and qualitative comparisons and ablation experiments are conducted to validate our method. M2HF has achieved the state-of-the-art performance for TVR on MSR-VTT, MSVD, LSMDC, DiDeMo, and ActivtiyNet. 


\end{document}